\documentclass[paper]{JHEP3} 

\usepackage{epsfig,multicol,amsmath,amsfonts,amssymb,cite}
\newcommand{\be}{\begin{equation}}
\newcommand{\ee}{\end{equation}}

\title{Qualitative solution of QCD sum rules}

\author{ S.S.~Afonin and  D.~Espriu\\
 Departament d'Estructura i Constituents de la Mat\`eria and
 CER for Astrophysics, Particle Physics and Cosmology,
 Universitat de Barcelona, 647 Diagonal, 08028, Spain\\
E-mail: \email{afonin@ecm.ub.es},
\email{espriu@ecm.ub.es}}

\preprint{February 2006\\UB-ECM-PF-06/08}

\abstract{We show how such important features of QCD as chiral
symmetry breaking or the formation of a mass-gap can be directly
traced from QCD sum rules for two point functions assuming, in the 
large number of colors limit, exact 
duality between the operator product expansion and the spectrum 
described by linearly (or nearly linear) rising Regge trajectories 
as predicted by string theory. We see how the presence of chiral 
symmetry breaking is intimately related to confinement in this scenario, as 
expected from general arguments, and how Regge trajectories 
change when chiral symmetry is broken.
As a result the whole meson mass
spectrum can be parametrized with a good accuracy by the constant
$f_{\pi}$ only, thus realizing the program proposed by Migdal some time ago.}


\begin{document}

\section{Introduction}

In the present work we consider QCD sum rules for the vector (V),
axial-vector (A), scalar (S), and pseudoscalar (P) channels in
the large-$N_c$ limit, the planar limit of QCD~\cite{hoof} and match the
corresponding operator product expansions (OPE) with the predictions of
 the QCD effective 
string theory. This approach has been used by several authors previously, (see
e.g.~\cite{we,all,last,afon,shif,ar}). In a sense, this work
may be partly viewed as a simplified version of the
analysis performed in~\cite{we} but with a rather different accent.

First of all, we show how the parameters of the effective string mass
spectrum can be extracted from the sum rules for the two-point functions 
in a very simple
way and without involved numerical fits (this is the main
distinction with~\cite{we}). Second, we try to give a clear
physical interpretation of all results appearing in the analysis.
We see how in this framework chiral symmetry breaking and confinement 
(i.e. the creation of a mass gap) are intimately related. Finally we provide in 
this rather well defined theoretical setting an implementation of Migdal's 
program whereby all physical quantities in theories like QCD are expressed, 
modulo simple numerical factors, in terms of a basic scale. $f_\pi$ is this
scale in the present case.   

The paper is organized as follows. In section~2 we introduce the
relevant formulas. Sections~3 and~4 are devoted to the analysis of linear
string-like mass spectrum for vector and scalar channels correspondingly. 
In sections~5 and~6 this analysis is extended
to the non-linear spectrum. The results obtained are
interpreted in section~7.
Section 8 is devoted to a comparison with previous results.
Final section~9 contains some concluding remarks.

\section{General formalism}

Let us briefly remind the reader of the general formalism for deriving 
QCD sum rules in the large-$N_c$ limit. In this limit the 
two-point correlation functions of quark currrents are saturated by an 
infinite set of narrow meson states with the quantum numbers of 
these currents, i.e. they can be represented in Euclidean space as follows
\begin{equation}
\label{cor1}
\Pi^J(Q^2)=\int d^4x\,e^{iQx}\langle\bar{q}\Gamma q(x)\bar{q}\Gamma q(0)
\rangle_{\text{planar}}=
\sum_n\frac{Z_J(n)}{Q^2+m_J^2(n)}+D_0^J+D_1^JQ^2,
\end{equation}
\begin{equation}
J\equiv S,P,V,A; \qquad
\Gamma=i,\gamma_5,\gamma_{\mu},\gamma_{\mu}\gamma_{5}; \qquad
D_{0},D_{1}=\text{const}.
\end{equation}
For simplicity we do not write the Pauli matrices but the $SU_f(2)$ 
symmetry is understood.
The last two terms are required for renormalization of infinite sums.
On the other hand, their high-energy asymptotics are given by 
the OPE~\cite{svz,rry,jam} (we consider the chiral limit)
\begin{equation}
\label{V}
\Pi^{V,A}(Q^2)=\frac{N_c}{12\pi^2}
\ln\!\frac{\Lambda^2}{Q^2}
+\frac{\alpha_s}{12\pi}\cdot
\frac{\langle G^2\rangle}{Q^4}+\frac{4\xi^{V,A}}{9}\pi\alpha_s
\frac{\langle\bar{q}q\rangle^2}{Q^6},
\end{equation}
\begin{equation}
\label{S}
\Pi^{S,P}(Q^2)=-\frac{N_c}{8\pi^2}Q^2
\ln\!\frac{\Lambda^2}{Q^2}
+\frac{\alpha_s}{8\pi}\cdot
\frac{\langle G^2\rangle}{Q^2}-\frac{2\xi^{S,P}}{3}\pi\alpha_s
\frac{\langle\bar{q}q\rangle^2}{Q^4},
\end{equation}
where
\begin{equation}
\xi^{V,P}=-7,\qquad \xi^{A,S}=11,
\end{equation}
and we have defined
\begin{equation}
\label{trans}
\Pi_{\mu\nu}^{V,A}(Q^2)\equiv\left(-\delta_{\mu\nu}Q^2+Q_{\mu}Q_{\nu}\right)
\Pi^{V,A}(Q^2).
\end{equation}
The symbols $\langle G^2\rangle$ and $\langle\bar{q}q\rangle$ denote the 
gluon and quark condensate, respectively.
The residues are parametrized as follows
\begin{equation}
\label{resid}
Z_{V,A}(n)\equiv2F_{V,A}^2(n),
\qquad Z_{S,P}(n)\equiv2G_{S,P}^2(n)m_{S,P}^2(n).
\end{equation}
with $F_{V,A}(n)$ being electromagnetic decay constants and $G_P(n)$ are 
related with the corresponding weak decay constants 
(see~\cite{we} for details).

The sum rules simply follow from comparison at each power of $Q^{-2}$ of 
the OPE~\eqref{V} and~\eqref{S} with the sum in Eq.~\eqref{cor1} after 
summing up over resonances (in a chiral invariant way)
and subtracting infinite constants which are irrelevant for our purposes.

\section{Sum rules for linear spectrum: Vector case}

Phenomenology tells us that the linear mass spectrum 
predicted by effective string theories (such as the ones represented by the 
Veneziano amplitude, the supersymmetric string, or the Lovelace-Shapiro 
amplitude) is a very good approximation to the real world~\cite{ani}. 
In the present section we shall adhere to this type of spectrum, 
namely assume that the spectrum is strictly linear and see 
what are the implications of this assumption. Thus
consider the ansatz
\begin{equation}
\label{lin}
m^2_J(n)=m^2+an, \qquad F^2(n)=\text{const}, \qquad n=0,1,2,\dots
\end{equation}
Here the universal slope $a$ is proportional to the string 
tension $T$: $a=2\pi T$.
After renormalization (subtraction of infinite constant) and taking 
into account the 
$\pi$-meson contribution (or more generally, the possibility 
of a zero-mass state in the pseudoscalar and axial-vector channels) we have
\begin{equation}
\label{V2}
\Pi^{V,A}(Q^2)=\frac{2f_{\pi}^2}{Q^2}\delta_{AJ}+
\sum_{n=0}^{\infty}\frac{2F_{V,A}^2}{Q^2+m^2+an}=
\frac{2f_{\pi}^2}{Q^2}\delta_{AJ}-\frac{2F_{V,A}^2}{a}\,\psi\!\left(
\frac{Q^2+m^2}{a}\right),
\end{equation}
where we have used the Kronecker symbol $\delta_{ab}$ to indicate that 
the hypothetical Goldstone boson contributes only to the axial-vector 
channel due to PCAC.

The $\psi$-function has an asymptotic representation at $z\!\gg\!1$
\be
\label{asymp}
\psi(z)=\ln{z}-\frac{1}{2z}-\sum_{k=1}^{\infty}\frac{B_{2k}}{2kz^{2k}},
\ee
there $B_{2k}$ denote the Bernoulli numbers.

Let us introduce the notation
\be
x\equiv\dfrac{m^2}{a}.
\ee
Performing the procedure outlined above one arrives at the following sum rules
\begin{align}
\label{1v}
\frac{N_c}{24\pi^2}&=\frac{F_J^2}{a}, \\
\label{2v}
0&=f_{\pi}^2\delta_{AJ}-F_J^2\left(x-1/2\right),\\
\label{3v}
\frac{\alpha_s}{12\pi}\langle G^2\rangle&=
F_J^2a\left(x^2-x+1/6\right),\\
\label{4v}
\frac{4\xi^{J}}{9}\pi\alpha_s\langle\bar{q}q\rangle^2&=
-\frac23F_J^2a^2x\left(x-1/2\right)\left(x-1\right).
\end{align}

From Eq.~\eqref{1v}
it follows that the quantities $F_J^2$
are directly expressed through the slope $a$. 
Let us substitute them into Eq.~\eqref{4v}.
The result is
\be
\label{vs}
-\frac{\xi^{J}}{N_c}\pi\alpha_s16\pi^2\frac{\langle\bar{q}q\rangle^2}{a^3}=
x\left(x-1/2\right)\left(x-1\right).
\ee

Now we note that the quantity 
$\langle\bar{q}q\rangle^2/a^3\approx(220/1100)^6\sim10^{-4}$ is very small 
in the real world. Thus the theory has a natural small parameter. 
More precisely, since the quantities $\xi^J/3$ and $\pi\alpha_s$ are of 
order one, the small parameter is
\be
16\pi^2\frac{\langle\bar{q}q\rangle^2}{a^3}\approx 0.01.
\ee
The existence of this small parameter in the sum rules seems to be 
tightly related to that of the chiral perturbation theory, i.e. 
the ratio squared of $m_{\pi}$ to the masses of light mesons in other channels.
A kind of "perturbation theory" can be developed here. In the zero order 
the sum rules become extremely simple providing nevertheless a 
good approximation to the real world. 
In the rest of the paper we will consider only this case.

In this approximation Eq. \eqref{vs} takes the form
\be
\label{vs1}
x\left(x-1/2\right)\left(x-1\right)\simeq 0,
\ee
and only two dimensionful parameters appear:
The slope $a$ which parametrizes the strength of the gluon interaction 
and, hence, must be directly related to the gluon 
condensate $\langle G^2\rangle$ (indeed we see that from (\ref{3v})),
and the $\pi$-meson weak decay constant $f_{\pi}$ parametrizing 
 chiral symmetry breaking\footnote{Throughout the paper we use 
the value of $f_{\pi}$ in the chiral limit~\cite{gasl}, $f_{\pi}=87$ MeV.}.

Let us investigate the solutions of the above set of equations.
First of all, equations \eqref{1v}-\eqref{3v} and \eqref{vs1} have 
one trivial solution (even without any approximations) corresponding  
to the case where the theory is weakly coupled (non-confining), 
without bound states and condensates: $F^2_J= 
a=0$ but the ratio in
Eq.~\eqref{1v} 
is non-zero and give the parton-model result. 
All non-perturbative effects are absent in this case.

Let us now analyze the possibility of a 
nontrivial\footnote{I.e. with a mass gap.} chirally 
symmetric solution with $f_{\pi}=0$ (here of course $\langle\bar{q}q\rangle$=0 
and Eq.~\eqref{vs1} is strictly true). From Eq.~\eqref{2v} we have 
only one possibility, namely $x=1/2$, which is consistent with Eq.~\eqref{vs1}.
Hence the chirally symmetric linear spectrum would be
\be
\label{ls}
m_J^2(n)=\frac{a}{2}+an.
\ee
Substituting the solution in Eq. \eqref{3v} and making use of 
Eq. \eqref{1v} we obtain a relation between the 
slope and the gluon condensate
\be
\label{G}
\frac{\alpha_s}{\pi}\langle G^2\rangle = -\frac{N_c a^2}{24\pi^2},
\ee
numerically resulting in
$\frac{\alpha_s}{\pi}\langle G^2\rangle\approx -(370\,\text{MeV})^4$ 
that is in agreement with the standard value in QCD,
$\frac{\alpha_s}{\pi}\langle G^2\rangle=(360\pm20\,\text{MeV})^4$, 
in absolute value, but not in the sign, which is an obvious inconsistency. 

In fact the gluon condensate is proportional to the QCD vacuum energy
$\varepsilon_{\text{vac}}$
(see discussions in section~7). As QCD is an asymptotic free theory the 
vacuum energy turns out to be negative for the stable vacuum~\cite{instab}. 
We draw attention to the quite interesting fact that the solution $x=1/2$ is 
nothing but the exact minimum of the 
function\footnote{The vacuum energy (or gluon condensate) in our 
approach turns out to be the function of intercept, which is quantized, 
i.e. this energy is also quantized.} 
$\varepsilon_{\text{vac}}\sim-\langle G^2\rangle\sim x^2-x+1/6$. Hence, 
in the vector channels the 
chirally symmetric case for the linear spectrum corresponds to the 
theory near the maximum of this function (since the sign is opposite
to that of vacuum energy).

Consequently, the chirally symmetric solution we have just been exploring 
cannot exist for the linear spectrum if confinement holds. 
Thus, the property of confinement 
in the large-$N_c$ limit automatically results in Chiral Symmetry Breaking (CSB), 
i.e. $f_{\pi}\neq0$. 
This agrees with the Coleman-Witten theorem~\cite{cw}: if confinement 
persists in the planar limit of QCD, chiral symmetry is inevitably broken.
As we will see later, the analysis of the chirally broken case necessarily
requires the introduction of non-linear corrections to the spectrum in 
any case.

Staying within the framework of the linear spectrum, we note that 
the chirally symmetric spectrum 
\eqref{ls} for the case of $\rho$-mesons agrees 
with the one derived from the so-called Lovelace-Shapiro (LS)
amplitude~\cite{LS} for the reaction $\pi+\pi\rightarrow\pi+\pi$,
 which had a great
phenomenological success in the 60's
\be
A(s,t)\sim\frac{\Gamma(1-\alpha_\rho(s)) \Gamma(1-\alpha_\rho(t))}
{\Gamma(1-\alpha_\rho(s)-\alpha_\rho(t))},
\ee
with $\alpha_\rho(s)=\frac{1}{2}+ \frac{s}{a}$.
This amplitude predicts 
equal masses for the ground
vector and scalar mesons, i.e. the solution $x=1/2$ would extend (if one were
to believe the LS amplitude) to both cases. An important property of this 
amplitude is the presence of an 'Adler zero'~\cite{adler}, i.e. the fact 
that $A(s,t)\to 0$ as $s,t \to 0$. This is actually a consequence of 
Goldstone theorem so there is some logical uneasiness here since we 
have assumed so far that chiral symmetry is unbroken, but being the presence
of the Adler zero a necessary but not sufficient condition for CSB one 
can find a scape route from contradiction. 

The generalization of this LS
amplitude to reactions of the form $\pi+A\rightarrow B+C$ was performed in~\cite{avw}.
The generalized LS amplitude describes the axial-vector 
and pseudoscalar states.
It turns out that the amplitude derived in \cite{avw} predicts 
$x=1$ for axial-vector 
and pseudoscalar mesons. This is numerically inconsistent with  
the sum rules (even if we include $f_{\pi}$) as they stand 
and, most importantly,  with chiral 
symmetry restoration at
high energies (see e.g. \cite{we}). The conclusion is that LS, in spite 
of its undeniable phenomeno
logical success, is unable to reproduce the sum rules
for two point functions. This rules it out as a possible model for QCD 
even leaving aside the fact that it does not derive from any known string 
theory. 

Coming back to Eqs.~\eqref{1v}-\eqref{4v} and
having ruled out the possibility of a chirally symmetric solution to 
the sum rules we have to look for solutions with $f_\pi \neq 0$. We know 
from current algebra that
this implies a non-zero value for the quark condensate. However, as we 
have discussed previously, the numerical value for the physically 
relevant quantity  $16\pi^2\langle\bar{q}q\rangle^2/a^3$ is very small,
so we can still use the approximate equation $x(x-
\frac{1}{2})(x-1)\simeq0$. If $x=\frac{1}{2}$ is ruled out because it 
corresponds to the chirally symmetric solution, can we perhaps consider
the other two solutions $x=0$ or $x=1$? The first one is obviously ruled
out due to sign inconsistencies. The second one is viable, but inconsistent 
with the sum rules unless we accept deviations from strictly linear 
trajectories. As we do not know how CSB distorts the linear spectrum, 
we will introduce these effects 
phenomenologically through non-linear corrections to the string-like spectrum.
Obviously chiral symmetry restoration at high energies requires
adopting the same value of $x$ in all channels~\cite{we,shif} and we will consider 
non-linear corrections in sections~5 and~6.

\section{Sum rules for linear spectrum: Scalar case}

It is convenient to make the rearrangement~\cite{we}:
\begin{multline}
\label{ren}
\Pi(Q^2)= 2\sum_n\frac{G^2(n)m^2(n)}{Q^2+m^2(n)} + D_0 +D_1 Q^2\\
=\left[\sum_n 2 G^2(n) + D_0\right] -
Q^2 \left[\sum_n\frac{2 G^2(n)}{Q^2+m^2(n)} - D_1\right],
\end{multline}
where the infinite constant $D_0$ and $D_1$ subtract the infinities 
coming from the sums over resonances.
After renormalization we have
\begin{equation}
\label{S2}
\Pi^{S,P}(Q^2)=\frac{Z_{\pi}}{Q^2}\delta_{PJ}-Q^2
\sum_{n=0}^{\infty}\frac{2G_{S,P}^2}{Q^2+m^2+an}=
\frac{Z_{\pi}}{Q^2}\delta_{PJ}+\frac{2G_{S,P}^2Q^2}{a}\,\psi\!\left(
\frac{Q^2+m^2}{a}\right).
\end{equation}
Let us consider again the chirally symmetric case. 
The pion residue $Z_{\pi}$ vanishes in this case and it
will be taken into account in section~6 where the chirally broken 
case is considered.
Performing the standard procedure (just as in the vector channels 
using the same notation $x$) one obtains the sum rules
\begin{align}
\label{1s}
\frac{N_c}{16\pi^2}&=\frac{G_J^2}{a}, \\
\label{2s}
\frac{\alpha_s}{8\pi}\langle G^2\rangle&=
-G_J^2a\left(x^2-x+1/6\right)
,\\
\label{3s}
-\frac{2\xi^{J}}{3}\pi\alpha_s\langle\bar{q}q\rangle^2&=
\frac23G_J^2a^2x\left(x-1/2\right)\left(x-1\right).
\end{align}

The general requirement
for the spectrum is the chiral symmetry restoration 
at high energies (i.e. equal $x$ for both channels). There are three 
solutions with these properties: $x=0,\,1/2,\,1$.
Substituting these solutions in~Eq. \eqref{2s} and making use 
of Eq. \eqref{1s} we obtain the relation between the slope 
and the gluon condensate
\be
\label{Gs}
\frac{\alpha_s}{\pi}\langle G^2\rangle = \frac{N_ca^2}{24\pi^2},
\qquad\text{for}\,\,x=\frac12,
\ee
\be
\label{Gs2}
\frac{\alpha_s}{\pi}\langle G^2\rangle = -\frac{N_ca^2}{12\pi^2},
\qquad\text{for}\,\,x=0\,,1.
\ee
The first case is in agreement with the spectrum of the LS amplitude 
for the scalar channel (but not for the pseudoscalar channel)
and the gluon condensate is reproduced with the correct sign this time
(see discussion after Eq.~\eqref{G}). The lightest pseudoscalar has the mass,
$m_{\pi}^2=a/2$, in agreement with the well known fact that if no 
spontaneous CSB occurs, then pion stays massive.
The appearance of the other two solutions is a consequence of the fact
that an analogue of Eq.~\eqref{2v} is absent in the scalar channels. 
The vacuum energy is evidently larger on these solutions, hence, 
they are unphysical. It is interesting to note that the gapless
solution that appears is among the energetically unfavorable ones. 
The existence of
this solution expresses another exact result: If a rigorously
chirally symmetric theory possesses a massless pseudoscalar bound state
then there is a degenerate scalar meson partner (see, e.g.,~\cite{holl}).            
The fact however that the value of $\langle G^2\rangle$ disagrees in sign
with Eq.~\eqref{G} is nevertheless another sign of the inconsistency
of the chirally unbroken solution.

\section{Sum rules for non-linear spectrum: Vector case}

According to the discussions at the end of section~3 we are 
forced to introduce a non-linear spectrum.
We will derive the minimal non-linear realization of the spectrum 
consistent with the sum rules.
For that we have to make some hypotheses about what represents the 
leading contribution in the non-linear case. First of all we know nothing 
from experiment about the residues of radially excited mesons. 
Numerical calculations show that they differ negligibly from the 
linear case~\cite{we}.
Thus for simplicity we do not introduce non-linear corrections for 
the residues of excited states and for $a_1$-meson as well 
(below we show that $F_{a_1}$ nicely agrees with experiment 
without any corrections). So, except for the $\rho$-meson, we denote
\be
\label{F}
F_V^2=F_A^2\equiv F^2.
\ee
We note also that the mass of $\rho$-meson is in the real world 
substantially below 
1 GeV, i.e. the typical CSB scale. Consequently it is natural to expect 
that this meson is affected by CSB to a larger extent than other 
vector mesons. We shall make the hypothesis 
that CSB induces a shift in the residue of the $\rho$ meson (and in no other 
vector or axial mesons). We shall later attempt to justify this
hypothesis. Then
\be
\label{tF}
F_{\rho}^2=F^2+\tilde{F}^2.
\ee

Now let us introduce the non-linear mass spectrum in the form
\be
m_J^2(n)=m^2+an+\delta_J(n),
\ee
where the corrections always can be represented as follows:
\be
\label{cor}
\delta_J(n)=aA_J f_J(n), \qquad f_J(0)=1.
\ee
Here the dimensionless constants $A_J$ are supposed to be due to CSB, 
i.e. they are proportional to $f_{\pi}^2/a$, and $f_J(n)$ are decreasing 
functions of $n$ only. 
They have to vanish at least exponentially in $n$ according to~\cite{we}, 
or polynomially according to~\cite{shif} (see discussion in section~8), 
but in this work we will not adhere to any particular model for these 
functions. If the non-linear corrections provide a rather 
big contribution to the masses of ground states but relatively 
small one to those of excited states (as we expect) then the 
parametrization \eqref{cor} automatically will not be very sensitive 
to a concrete choice of $f_J(n)$.

An important assumption of the present analysis is that the slope $a$ has 
no corrections. This is a universal quantity which does not feel CSB. Any 
corrections to the spectrum are encoded in the value of intercept $m^2$ and
in the non-linear contribution $\delta(n)$. Such a picture is well justified 
phenomenologically.

The contribution of the non-linear mass corrections to the correlator 
is (for the time being we denote $\delta\equiv\delta_J(n)$ and omit 
the general factor)
\begin{multline}
\frac{1}{Q^2+m^2+an+\delta}=\frac{1}{Q^2+m^2+an}
\left(1+\frac{\delta}{Q^2+m^2+an}\right)^{-1}=\\
\frac{1}{Q^2+m^2+an}-\frac{\delta}{Q^4}+\frac{1}{Q^6}
\left(2m^2\delta+2an\delta+\delta^2\right)+\mathcal{O}
\left(\frac{1}{Q^8} \right) .
\label{dec}
\end{multline}
The $\rho$-meson additional contribution is
\be
\frac{\tilde{F}^2}{Q^2+m^2+aA_V}=
\frac{\tilde{F}^2}{Q^2}-\frac{\tilde{F}^2a(x+A_V)}{Q^4}+
\frac{\tilde{F}^2a^2(x+A_V)^2}{Q^6}+\mathcal{O}\left(\frac{1}{Q^8}\right).
\ee
Let us introduce the constants
\be
\label{cs}
C_0^J\equiv\sum_{n=0}^{\infty}f_J(n),\qquad C_1^J
\equiv\sum_{n=0}^{\infty}nf_J(n),\qquad C_2^J\equiv\sum_{n=0}^{\infty}f^2_J(n).
\ee
After summation over $n$ and making use of notations \eqref{F}, 
\eqref{tF}, and \eqref{cs} sum rules \eqref{1v}-\eqref{4v} become
\be
\label{1}
\frac{N_c}{24\pi^2}=\frac{F^2}{a},
\ee
\be
\label{2}
f_{\pi}^2\delta_{AJ}+\tilde{F}^2\delta_{VJ}=F^2\left(x-1/2\right),
\ee
\be
\label{3}
\frac{\alpha_s}{12\pi F^2a}\langle G^2\rangle=
x^2-x+1/6-2\Delta_1^J,
\ee
\be
\label{4}
0=x\left(x-1/2\right)\left(x-1\right)-3\Delta_2^J,
\ee
where the corrections $\Delta_i^J$ are non-zero due to CSB 
and they are given by
\begin{gather}
\Delta_1^J=A_JC_0^J+\frac{\tilde{F}^2}{F^2}(x+A_V)\delta_{VJ},\\
\Delta_2^J=A_J\left(2C_0^Jx+2C_1^J+C_2^JA_J\right)+
\frac{\tilde{F}^2}{F^2}(x+A_V)^2\delta_{VJ}.
\end{gather}
From Eq. \eqref{2} we get
\be
\label{tilF}
\tilde{F}=f_{\pi}.
\ee
Thus from the same equation one has
\be
\label{x}
x=\frac12+\frac{f_{\pi}^2}{F^2}=\frac12+\frac{24\pi^2f_{\pi}^2}{N_ca}.
\ee
At this stage we need some additional input to advance. For reliability
this input should be a quantity which is experimentally known with 
a good accuracy. Except $f_{\pi}$ (which is already an input) the best 
candidate for such a quantity
is $F_{\rho}=154\pm8$ MeV. We note now that this number is very close to
$\sqrt{3}\,f_{\pi}\approx151$ MeV. Thus, within our accuracy, we may put
$F_{\rho}=\sqrt{3}\,f_{\pi}$ as input. Making use of Eqs.~\eqref{F} 
and~\eqref{tilF} one obtains 
\begin{equation}
F=\sqrt{2}\,f_{\pi}\approx123\,\text{MeV},  
\end{equation}
which is in agreement with the experimental value for the decay constant
of $a_1$-meson (see~\eqref{F}): $F_{a_1}=123\pm25$ MeV. From Eq.~\eqref{x}
we immediately get the solution for the intercept, $x=1$, and the relation 
\be
\label{fund}
a=\frac{48}{N_c}\pi^2f_{\pi}^2,
\ee
in very good agreement with phenomenology. This formula happens to be
the result reported in~\cite{T} based on completely different grounds.
It is curious to note that relation~\eqref{fund} can be obtained 
from combining 
the slope of the LS amplitude, $a=2m_{\rho}^2$ (we remind that
the slope has no corrections in the present analysis), with a known 
formula for the mass of $\rho$-meson~\eqref{VMD} (both relations
were also found in~\cite{afon}). In principle, Eq.~\eqref{fund} 
could be taken as input from the very beginning, all other results
are then automatically reproduced. This way would be natural in the 
sense that considering the CSB case we should insert a spectrum of string 
with CSB incorporated. Such strings were considered, e.g., in~\cite{T2}.
Further development of this approach resulted in a qualitative
derivation of Eq.~\eqref{fund} in~\cite{T},
where the scales of the Goldstone boson physics and of the string 
dynamics are supposedly related through Eq.~\eqref{fund} by the chiral anomaly.

It is thus quite remarkable that CSB forces a jump from $x=\frac12$
to $x\simeq1$. A physical interpretation of the intercept is that
it is a typical mass of ground state, hence, it should be tightly related
to the scale of CSB, $\Lambda_{\text{CSB}}\simeq4\pi f_{\pi}$. 
Accidentally or not, they are identical, 
$m^2\simeq a\simeq\Lambda_{\text{CSB}}^2$. A deeper understanding of this
coincidence might lead to an independent derivation of Eq.~\eqref{fund}.

From the last sum rule Eq. \eqref{4} one can see that at the 
solution $x=1$ the contribution due to the linear part of the spectrum 
(the first term in the r.h.s.) is exactly cancelled. As a result the 
corrections turn out to be directly related to the (unwritten) quark 
condensate in the l.h.s. This is in a full agreement with our initial 
assumption concerning the nature of the corrections.

The slope $a$ has a purely gluonic nature and, consequently, its value 
should not change dramatically after CSB. Thus, one may substitute 
Eq.~\eqref{fund} 
into Eq.~\eqref{ls} and obtain the mass of ground state in the chirally
symmetric case: 
\be
\label{VMD}
m^2_J(0)=\frac{24}{N_c}\pi^2f_{\pi}^2.
\ee 
This is nothing but a formula
for the mass of $\rho$-meson obtained some time ago within the 
Borel QCD sum rules~\cite{mrho} and rederived in some 
other types of sum rules (see, e.g.,~\cite{afon,weise}). This 
formula also usually holds when matching different models to the 
vector meson dominance and appears all over the literature
(see, e.g.,~\cite{ar2,son1}). Relation~\eqref{VMD} reproduces 
the experimental mass of $\rho$-meson with a unexpectedly high accuracy
(we remind that $f_{\pi}=87$~MeV in the chiral limit). Thus,
we should conclude that the mass of the ground vector meson practically 
does not shift after CSB. Then we immediately obtain the constant $A_V$
parametrizing corrections in the vector channel
\be
\label{Av}
A_V=-\frac{24\pi^2f_{\pi}^2}{N_ca}=-\frac12.
\ee
Substituting Eqs. \eqref{tilF}, \eqref{Av}, and \eqref{Gs} 
(which can be rewritten as $\frac{\alpha_s}{\pi}\langle G^2\rangle=F^2a$) 
into Eq. \eqref{3} we obtain the estimate
\be
\label{Cv}
C_0^V=\frac{5}{12}.
\ee
From the same equation for the axial-vector channel we have the relation
\be
A_AC_0^A=\frac{1}{24}.
\ee
For a rough estimate of the $a_1$-meson mass we may accept a 
kind of universality,
$C_0^A\approx C_0^V$, which leads to 
$m_{a_1}^2\approx1.1a\approx(1150\,\text{MeV})^2$.
Exactly the same value was obtained in the framework of QCD sum rules~\cite{rry}.
It seems that a better estimate is hardly possible within the sum rules,
let alone the fact that in the present work we have accepted the large-$N_c$ limit. 
This example shows that the mass corrections are indeed 
relatively small for the mesons with the mass above the CSB scale. 
Even for the $a_1$-meson they constitute approximately 10\% which competes with the accuracy of the large-$N_c$ counting. This is what we expected from the very beginning. Keeping in mind that the constants $C_i^J$ are of order 1 thanks to the assumed fast convergence of functions $f_J(n)$, we could independently recover in this way the approximate solution $x=1$ in the CSB phase from Eq. \eqref{4} for the axial-vector case (knowing from Eq. \eqref{2} that $x=0$ and $x=1/2$ are not solutions).

\section{Sum rules for non-linear spectrum: Scalar case}

As in the previous section we will be interested in the minimal non-linear realization of the spectrum consistent with the sum rules.
Thus, we do not introduce non-linear corrections for the residues of excited states. Let us denote
\be
\label{GG}
G_S^2=G_P^2\equiv G_0^2.
\ee
For the ground states parametrization~\eqref{resid} should be changed
\begin{equation}
Z_J(0)\longrightarrow Z_J(0)+\tilde{Z}_J.
\end{equation}
Here $\tilde{Z}_P$ is the $\pi$-meson residue which cannot be taken into account within parametrization~\eqref{resid} and $\tilde{Z}_S$ reflects a contribution to the residue of ground scalar state.
Introducing non-linear corrections to the masses in the form~\eqref{cor} and repeating the simple calculations outlined in the previous section one arrives at the sum rules
\begin{align}
\label{1sn}
\frac{N_c}{16\pi^2}&=\frac{G_0^2}{a}, \\
\label{2sn}
\frac{\alpha_s}{8\pi}\langle G^2\rangle&=
-G_0^2a\left(x^2-x+1/6\right)+\tilde{Z}_{J},\\
\label{3sn}
0&=x\left(x-1/2\right)\left(x-1\right)-3(x+A_J)\left(A_JC_0^J+\frac{\tilde{Z}_S\delta_{JS}}{2aG_0^2}\right).
\end{align}
Motivated by the previous results, we assume that the linear part of the spectrum does not
induce CSB.
The physical solutions of the system are then $x=1$ for the scalar channel and $x=1$ or $x=0$
for the pseudoscalar one. In both cases we obtain the relation
\begin{equation}
\label{Zps}
\tilde{Z}_{P}=\tilde{Z}_{S}=\frac{\alpha_s}{8\pi}\langle G^2\rangle+\frac{N_ca^2}{96\pi^2}.
\end{equation}
Using the current algebra result,
\be
\label{cural}
\tilde{Z}_{P}=2\frac{\langle\bar{q}q\rangle^2}{f_{\pi}^2},
\ee
and phenomenological values for the gluon condensate and the slope
one can estimate the quark condensate
\be
\langle\bar{q}q\rangle
\approx-(170\,\text{MeV})^3.
\ee
Taking into account our approximations, this estimation is selfconsistent. 
Combined Eqs.~\eqref{Zps} and~\eqref{cural} pressupose a certain relation between
the vacuum condensates and hadronic parameters, which is similar to a result in~\cite{afon3}.
For $x=1$ one has
$A_P=-1$ and the $n$-th pseudoscalar meson is a chiral partner for the $n$-th scalar meson.
In the second case $A_P=0$ and the $(n+1)$-th pseudoscalar meson is a chiral partner for
the $n$-th scalar meson. In this case there is no non-linear correction to the pseudoscalar
spectrum and the $\pi$-meson has no chiral partner.

For the scalar channel we have from Eqs.~\eqref{3sn} and~\eqref{Zps}
\be
A_SC_0^S=-\frac18.
\ee
In order to estimate the mass of the ground state we can, like in the axial-vector channel,
assume an approximate universality for the non-linear contributions, $C_0^S\approx C_0^V$. Making use
of Eq.~\eqref{Cv} this leads to $m_{S}^2(0)\approx0.7a\approx(920\,\text{MeV})^2$. The relative
contribution to the residue of ground state due to non-linear corrections is of order
$\tilde{Z}_{S}/Z_S(0)=0.1\div0.2$, i.e. much less than for the $\rho$-meson.

\section{Discussions}

Let us give an intuitive interpretation which lies behind the results obtained. First of all, the gluon condensate is a measure of vacuum energy
$\varepsilon_{\text{vac}}$. To be more precise
we remind that the vacuum energy and the nonperturbative part (n.p.) of v.e.v. $\langle G^2\rangle$ are related by the anomaly in the trace of energy-momentum tensor~\cite{anom}
\be
\label{anomaly}
4\varepsilon_{\text{vac}}=\langle \Theta_{\mu}^{\mu}\rangle_{\text{n.p.}}=\frac{\beta(\alpha_s)}{4\alpha_s}\langle G^2\rangle_{\text{n.p.}}+
\mathcal{O}(\alpha)+\cdots.
\ee
The term $\mathcal{O}(\alpha)$ is the contribution of quark polarization effects.
In~\cite{mig} the effective potential for $\Theta_{\mu}^{\mu}$ was constructed in the tree approximation.
It was shown that the minimum of this potential (vacuum energy) must be negative, i.e. as the $\beta$-function is negative only a positive gluon condensate provides a stable vacuum.

In our analysis it is instructive to regard mesons as the spectrum of excitations given
by the effective normalized quark vacuum "potential energy" $U$, where
\be
\label{U}
U\equiv-\frac{\alpha_s}{\pi}\frac{\langle G^2\rangle}{a/(4\pi^2)},
\ee
and the gluon condensate $\langle G^2\rangle$ depends on the vacuum energy
$\varepsilon_{\text{vac}}$.

Let us take the vector and axial-vector channels as an example and discuss the three solutions of the sum rules mentioned in the paper. The trivial solution corresponds to
the weakly-coupled QCD where the three- and four-point gluon interactions are suppressed and we effectively have $U_{\text{min}}=0$ like in the QED. Since the vacuum is perturbative there are neither condensates nor resonances.

The chirally-symmetric (CS) solution corresponds to fluctuations over the perturbative vacuum. The potential energy $U$ has a local maximum in this point, $U_{\text{max}}=m_{\rho}^2$, which results in a negative gluon condensate. As we still calculate the excitations over the zero-level, there appear the mass gap $\Delta_{\text{CS}}=m_{\rho}^2=a/2$.

After the CSB the potential energy recovers its minimal value, $U_{\text{min}}=-m_{\rho}^2$.
This new minimum turns out to be lower than the perturbative one by the value $8\pi^2f_{\pi}^2$ which is equal to $m_{\rho}^2$. As a result the mass-gap enlarges by two times,
$\Delta_{\text{CSB}}=2m_{\rho}^2=a=(4\pi f_{\pi})^2$.
It is exactly equal to the generally accepted value for the gap squared of spontaneous chiral symmetry breaking $\Lambda_{\text{CSB}}^2$!

Let us try to understand the "special" nature of $\rho$-meson in the framework of presented approach. In the perturbative vacuum we always have photons which are the only massless colorless particles in the Standard Model.
However, in the nonperturbative vacuum this vector particle "feels" the "deep" below $U=0$ through its virtual quark-antiquark components. As a result the quark and antiquark in the virtual pair can acquire a dynamical (constituent) mass, converting the photon into a massive vector particle. In other words, we can observe the photon over this deep as a vector particle with the mass $m_{\rho}$.
Since the height of hump over zero-level and that of deep below this level are equal (the perturbative vacuum is distorted symmetrically in the both directions), the $\rho$-meson mass is not shifted after the CSB. The described effect in a natural way results in the enhancement (compared to other mesons) of probability for creation of this meson from vacuum, i.e. the positive shift of $\rho$-meson residue. This interpretation provides an intuitive picture about the mechanism of vector meson dominance.

In the axial-vector channel the situation is completely different. The $a_1$-meson pole is shifted in the CSB phase supposedly due to the $\pi-a_1$ mixing, i.e. the $a_1$-meson acquires additional mass due to a very intensive interaction with the Goldstone bosons.
The mechanism seems to be the following. In the chirally symmetric phase both masses are equal,
$m_{a_1}^2=m_{\pi}^2=a/2$. In the CSB phase these particles become maximally mixed. If two particles are mixed,
rotation of their fields to a physical basis is known to result in repulsion of the initial masses. If the mixing
is maximal the repulsion is then also maximal and in our case leads to $m_{a_1}^2=a/2+a/2=a$,  
$m_{\pi}^2=a/2-a/2=0$. This interpretation is in a full agreement with old results of current algebra~\cite{gh,wein}.
Namely, if the chiral symmetry were unbroken the ground states would fall into irreducible representations
of $SU(2)\otimes SU(2)$ chiral group, with the masses being equal insight a given representation. After CSB these representations get mixed. In particular, the following relation holds: 
$m_{\pi}^2\cos^2\psi+m_{a_1}^2\sin^2\psi=m_{\rho}^2$, where $\psi$ is a mixing angle between pseudoscalar and
axial-vector representations. This angle was determined from the $\rho$ width to be around $45^{\text{o}}$, i.e. the mixing is maximal. In the chiral limit the relation converts then into the Weinberg formula. It is interesting to note that the scalar and vector representations practically do not mix after CSB. It is another manifestation of the fact that $\rho$-meson mass does not shift after CSB. So it does for the ground scalar meson. In practice, however, this state could have glueball, four-quark and strange components which hamper the prediction of its mass on the base of considered sum rules.  


Thus the underlying reason for the appearance of mass shifts 
in the CSB phase is the formation of constituent quarks which 
interact with the Goldstone bosons in a different way in channels 
with different quantum numbers. This effect is drastic for the 
ground states, the higher is the radial excitation the less 
important is the CSB phenomenon and the chiral symmetry gets 
restored~\cite{shif,gloz}.

Let us emphasize that the formulas like $m_{\rho}=\sqrt{8}\,\pi f_{\pi}$ 
make sense only in the CSB phase. This point sometimes causes a 
confusion in the literature (see e.g.~\cite{weise}). The mentioned relation 
by no means signifies that in-medium mass of $\rho$-meson changes in response 
to changing $f_{\pi}$, it can not be an order parameter for phase transitions 
like $f_{\pi}$. The primordial formula is $m_{\rho}=\sqrt{a/2}$, i.e. this 
mass has a gluonic origin. What we should expect at high temperatures or baryon 
densities is a decreasing $m_{a_1}$ up to the value $m_{\rho}$, the latter 
being unchanged in the first approximation. At the same time $m_{\pi}$ should 
increase up to the same value $m_{\rho}$. Finally, all masses of ground states 
become equal to $\sqrt{a/2}$ in the chirally restored phase.

\section{Relation to other work}

Concerning the nature of non-linear corrections to the string-like spectrum
we would like to mention an interesting paper by Weinberg~\cite{wein}. In that 
approach any mass matrix can be written as the
sum of chirally invariant term and chirally non-invariant one. The 
latter term appears due to a non-trivial matrix of axial couplings and
destroys the chiral symmetry in the spectrum. Our analysis suggests a one-to-one 
correspondence between the chirally invariant linear part of the spectrum
and the first term, and the non-linear contribution and the second term in~\cite{wein}.
For the higher excitations the matrix of axial couplings is expected to
gradually get trivial~\cite{afon2} resulting in suppression of the second
term. This scenario realizes an expectation (see, e.g., discussions in~\cite{GN}
and references therein) that restoration of chiral symmetry in the hadron 
spectrum is related to decoupling of Goldstone bosons from highly excited 
states.
 
In some models based on sum rules at large-$N_c$~\cite{last,afon,ar} the 
intercepts in spectrum of chiral partners are different. This 
discrepancy with our results (and those of~\cite{we,shif}) can be confusing.
We note, however, that such type of spectra are usually obtained using
some additional assumptions. In~\cite{ar} both models were designed with 
the aim to obtain a non-zero dimension-two gluon condensate. The price to pay
is, e.g., that the well-motivated relation~\eqref{fund} does not hold. 
In the model in~\cite{last} the expression for low-energy constant $L_8$ was imposed
as "another one" sum rule and then fine-tuned by a special prescription
for the cut-off in the infinite sums. As a result this constant comes from
perturbative logarithm, i.e., it is identically zero for any finite
number of resonances (lack of a smooth transition from finite number of states
to infinite one). The ideas in~\cite{afon} are somewhat close to those of
present paper, but in that paper there is no natural transition to the chirally 
symmetric case and the chosen way to introduce the non-linear corrections is
rather questionable.

{\bf 
In the literature exist two different estimations for the rate of chiral
symmetry restoration. In~\cite{shif} the power-law minimal fall-off was obtained
from matching to OPE without any convergence requirements for the sum rules,
while in~\cite{we} the fall-off was estimated to be exponential after imposing
the convergence for the generalized Weinberg sum rules. At present it is difficult
to say which variant is realized in nature.
}


Finally we would like to mention a possible relation of our work to the so-called
AdS/QCD models which are believed to provide a precise correspondence between
conformal gauge theories (like QCD in the ultraviolet 
region) and some superstring theories (see, e.g.,~\cite{kleb} and references 
therein). These models can naturally incorporate chiral properties of QCD,
in particular, the Gell-Mann-Oakes-Renner formula was derived in~\cite{son2}
and relation to some effective models with hidden gauge bosons was 
demonstrated in~\cite{son1}. As was noticed in~\cite{shif} and
developed in~\cite{low} they are often dual to the Migdal approach
to the QCD sum rules~\cite{prog} based on Pad\'e approximation
of the two-point correlators. Thus, QCD sum rules in the large-$N_c$
limit and AdS/QCD models seem to be complementary approaches. 
In this respect an interesting result was reported in~\cite{son3}.
Namely, at certain constraints on the infrared behavior of AdS/QCD
one can obtain the linear spectrum for the vector and axial mesons
with the same slope. The next successful step in this direction
would be a derivation of AdS/QCD model which reproduces equal
intercepts to be consistent with chiral symmetry restoration at high 
energies. Another challenge for the AdS/QCD approach would be
derivation of the relation for the string slope, Eq.~\eqref{fund}.

\section{Conclusions}

In the late 70's before the publication of 
Shifman-Vainstein-Zakharov paper~\cite{svz}
A.~Migdal formulated the program of calculation of the whole
spectrum for mesons made of light quarks as a function of some
unique dimensional parameter $\mu$: $M_i=\mu C_i$, where the
factor $C_i$ depends on the discrete input parameters of QCD such
as the number of colours and flavors~\cite{prog}. This formula was
supposed to be possible due to small values of bare quark masses
which can be neglected (except the Goldstone bosons, of course).

The approach presented in this paper is, in a sense, a concrete
realization of Migdal program.
Indeed, the spectrum depends on the constant~$f_{\pi}$, the slope
of trajectories, and the gluon condensate. These three parameters
are related by Eqs.~\eqref{Gs} and~\eqref{fund}. Consequently, we
may choose only one of them to parametrize the whole spectrum. The
most natural choice is evidently~$f_{\pi}$. The first correction
to the spectrum comes from taking into account the quark
condensate squared just as was supposed by Migdal in the same
paper! The extension of this analysis to mesons with other quantum
numbers and to the $SU_f(3)$ group is a rather technical work.

\acknowledgments

We are grateful to A.A.~Andrianov for fruitful discussions and enlightening comments.
The work was supported by CYT FPA, grant 2004-04582-C02-01, and by CIRIT GC, grant 2001SGR-00065.
The work of S.A. was also supported by RFBR, grant 05-02-17477.
S.A. thanks the Ministry of Education and Science of Spain for financial support.


\begin{thebibliography}{99}
\bibitem{hoof} G.~'t Hooft, \npb{72}{1974}{461};\\
E.~Witten, \npb{160}{1979}{57}.
\bibitem{we} S.S.~Afonin, A.A.~Andrianov, V.A.~Andrianov and D.~Espriu, \jhep{04}{2004}{039};\quad
\hepph{0403268}.
\bibitem{all} A.~Bramon, E.~Etim and M.~Greco, \plb{41}{1972}{609};\\
M.~Greco, \npb{63}{1973}{398};\\
J.J.~Sakurai,\plb{46}{1973}{207};\\
N.V.~Krasnikov and A.A.~Pivovarov, \yf{35}{1982}{1270};\quad
\plb{112}{1982}{397};\\
A.L.~Kataev, N.V.~Krasnikov and A.A.~Pivovarov,
\plb{123}{1983}{93};\\
S.G.~Gorishnii, A.L.~Kataev and S.A.~Larin,
\plb{135}{1984}{457};\\
E.A.~Tainov, \zpc{10}{1981}{87};\\
B.V.~Geshkenbein, \yf{42}{1985}{991};\\
B.V.~Geshkenbein,  \sjnp{49}{1989}{705};\\
M.~Shifman, {\it At the Frontier of Particle Physics/Handbook
of QCD,} (ed. M.~Shifman), World Scientific, 2001;\quad
\hepph{0009131};\\
M.~Golterman and S.~Peris, \jhep{01}{2001}{028};\\
S.R.~Beane, \prd{64}{2001}{116010};\\
S.R.~Beane, \plb{521}{2001}{47};\\
Yu.A.~Simonov, \pan{65}{2002}{135};\quad
\hepph{0109081};\\
M.~Golterman, S.~Peris, B.~Phily and E.~de~Rafael,
\jhep{01}{2002}{024};\\
A.A.~Andrianov, V.A.~Andrianov and S.S.~Afonin, \hepph{0212171};\\
M.~Golterman and S.~Peris, \prd{67}{2003}{096001};\\
V.I.~Shevchenko and Yu.A.~Simonov, \prd{70}{2004}{074012}; \quad \hepph{0406276};\\
A.L.~Kataev, \pan{68}{2005}{567};\quad \hepph{0406305};\\
O.~Cat\`a, M.~Golterman and S.~Peris, \jhep{05}{2005}{076};\\
J.J.~Sanz-Cillero, \npb{732}{2006}{136};\quad \hepph{0507186};\\
V.D.~Orlovsky, \hepph{0510192};\\
S.S.~Afonin, A.A.~Andrianov, V.A.~Andrianov and D.~Espriu, \ijmpa{21}{2006}{885}; 
\quad \hepph{0509144};\\
S.S.~Afonin, \hepph{0606291}.
\bibitem{last} O. Cat\`a, M. Golterman and S. Peris, \hepph{0602194}.
\bibitem{afon} S.S.~Afonin, \plb{576}{2003}{122}.
\bibitem{shif} M.~Shifman, \hepph{0507246}.
\bibitem{ar} E.R.~Arriola and W.~Broniowski, \hepph{0603263}.
\bibitem{svz} M.A.~Shifman, A.I.~Vainstein and V.I.~Zakharov,
\npb{147}{1979}{385, 448}.
\bibitem{rry} L.J.~Reinders, S.~Yazaki and H.R.~Rubinstein,
\npb{196}{1982}{125}.
\bibitem{jam} M.~Jamin and M.~M\"unz, \zpc{60}{1993}{569};\quad \hepph{9208201}.
\bibitem{ani} A.V.~Anisovich, V.V.~Anisovich and A.V.~Sarantsev,
\prd{62}{2000}{051502};\\
V.V.~Anisovich, \ufn{47}{2004}{49};\quad \hepph{0208123};\\
D.V.~Bugg, \prep{397}{2004}{257}.
\bibitem{gasl} J.~Gasser and H.~Leutwyler, \npb{250}{1985}{465}.
\bibitem{instab} M.A.~Shifman, A.I.~Vainstein and V.I.~Zakharov,
\jetpl{27}{1978}{55};\\
A.A.~Andrianov, V.A.~Andrianov, V.Yu.~Novozhilov and Yu.V.~Novozhilov,
\plb{186}{1987}{401};\\
A.A.~Andrianov and V.A.~Andrianov, \zpc{55}{1992}{435}.
\bibitem{cw} S.~Coleman and E.~Witten, \prl{45}{1980}{100}.
\bibitem{LS} C.~Lovelace, \plb{28}{1968}{264};\\
J.A.~Shapiro, \pr{179}{1969}{1345}.
\bibitem{adler} S.~Adler, \pr{137}{1965}{B1022}.
\bibitem{avw} M.~Ademollo, G.~Veneziano and S.~Weinberg,
\prl{22}{1969}{83}.
\bibitem{holl} A.~H\"oll, A.~Krassnigg and C.D.~Roberts, \prc{70}{2004}{042203}.
\bibitem{T} A.A.~Andrianov, D.~Espriu and A.~Prats, {\it AIP Conf.\ Proc.\ }{\bf 756} (2005) 302;
\quad \hepph{0412380}.
\bibitem{T2} J.~Alfaro, A.A.~Andrianov, L.~Balart and D.~Espriu, \ijmpa{18}{2003}{2501}.
\bibitem{mrho} M.A.~Shifman, A.I.~Vainstein and V.I.~Zakharov,
\prl{42}{1979}{297}.
\bibitem{weise} E.~Marco and W.~Weise, \plb{482}{2000}{87}.
\bibitem{ar2} E.R. Arriola and W.~Broniowski, \prd{67}{2003}{074021};\\
E.~Megias, E.R. Arriola, L.L.~Salcedo and W.~Broniowski, \prd{70}{2004}{034031}.
\bibitem{son1} D.~T.~Son and M.~A.~Stephanov, \prd{69}{2004}{065020}.
\bibitem{afon3} S.S.~Afonin, \hepph{0603176}.
\bibitem{anom} J.S.~Collins, L.~Duncan and S.D.~Joglekar, \prd{16}{1977}{438}.
\bibitem{mig} A.A.~Migdal and M.A.~Shifman, \plb{114}{1982}{445};\\
M.A.~Shifman, \prep{209}{1991}{341}.
\bibitem{gh} F.~J.~Gilman and H.~Harari, \pr{165}{1968}{1803}.
\bibitem{wein} S.~Weinberg, \pr{177}{1969}{2604}.
\bibitem{gloz} A.~Le~Yaouanc, L.~Oliver, S.~Ono, O.~P\`ene and J.-C.~Raynal, 
\prd{31}{1985}{137};\\
A.A.~Andrianov and V.A.~Andrianov, \hepph{9705364};\quad \hepph{9911383};\\
A.A.~Andrianov, D.~Espriu and R. Tarrach, \npb{533}{1998}{429};\\
A.A.~Andrianov and D.~Espriu, \jhep{10}{1999}{022};\\
L.Ya.~Glozman, \plb{475}{2000}{329};\\
A.A.~Andrianov, V.A.~Andrianov and S.S.~Afonin, \hepph{0101245};\quad \hepph{0209260};\\
V.A.~Andrianov and S.S.~Afonin, \yf{65}{2002}{1913};\quad \hepph{0109026};\\
V.A.~Andrianov and S.S.~Afonin, {\it Eur.\ Phys.\ J. }{\bf A 17} (2003) 111;\\
V.A.~Andrianov and S.S.~Afonin, {\it J.\ Math.\ Sc.\ }{\bf 125} (2005) 99;\quad
\hepph{0304140};\\
E.S.~Swanson, \plb{582}{2004}{167};\quad \hepph{0309296};\\
T. DeGrand, \prd{69}{2004}{074024};\\
L.Ya.~Glozman, A.V.~Nefediev and J.E.~Ribeiro, \prd{72}{2005}{094002};\\
L.Ya.~Glozman, \hepph{0411281};\\
S.S.~Afonin, \plb{639}{2006}{258};\quad \hepph{0603166};\\
S.S.~Afonin, \hepph{0606310}.
\bibitem{afon2} S.S.~Afonin, \hepph{0605102}.
\bibitem{GN} L.Ya.~Glozman and A.V.~Nefediev, \hepph{0603025}.
\bibitem{kleb} I.R. Klebanov, \hepph{0509087}.
\bibitem{son2} J.~Erlich, E.~Katz, D.T.~Son and M.A.~Stephanov, \prl{95}{2005}{261602}.
\bibitem{low} J.~Erlich, G.D.~Kribs and I.~Low, \prd{73}{2006}{096001}.
\bibitem{prog} A.A.~Migdal, \ap{110}{1978}{46}.
\bibitem{son3} A.~Karch, E.~Katz, D.T.~Son and M.A.~Stephanov, \hepph{0602229}.

\end{thebibliography}
\end{document}